\documentclass[floatfix,superscriptaddress,twocolumn,showpacs,prb,amsmath,amssymb]{revtex4} 
\usepackage{graphicx} % Include figure files

\newcommand{\apo}{Ag$_5$Pb$_2$O$_6$}
\newcommand{\sub}[1]{$_{\mathrm {#1}}$}
\newcommand{\subm}[1]{_{\mathrm {#1}}}
\newcommand{\etal}{\textit{et al.}}
\newcommand{\degc}{$^{\circ}$C}

\begin{document}

\title{Non-linear Temperature Dependence of Resistivity in Single Crystalline Ag$_5$Pb$_2$O$_6$}

\author{Shingo~Yonezawa}
\affiliation{Department of Physics, Graduate School of Science, 
Kyoto University, Kyoto, 606-8502, Japan}

\email{yonezawa@scphys.kyoto-u.ac.jp}

\author{Yoshiteru~Maeno}
\affiliation{Department of Physics, Graduate School of Science, 
Kyoto University, Kyoto, 606-8502, Japan}
\affiliation{International Innovation Center, 
Kyoto University, Kyoto, 606-8501, Japan}

\date{\today}

%.Abstract

\begin{abstract}
We measured electrical resistivity, specific heat and magnetic susceptibility
of single crystals of highly conductive oxide Ag$_5$Pb$_2$O$_6$,
which has a layered structure containing a Kagome lattice.
Both the out-of-plane and in-plane resistivity show $T^2$ dependence 
in an unusually wide range of temperatures up to room temperature.
This behavior cannot be accounted for either by electron correlation
or by electron-phonon scattering with high frequency optic phonons.
In addition, a phase transition with a large diamagnetic signal 
was found in the ac susceptibility,
which strongly suggests the existence of a superconducting phase below 48~mK.
\end{abstract}


\maketitle

\section{Introduction}

One of the central issues of solid state physics in recent decades has been 
the physics of electron-electron (e-e) correlation.
As a hallmark of the Fermi liquid state formed by strong electron correlations, 
the resistivity exhibits $T^2$ dependence at low temperatures,
as in many heavy-fermion systems\cite{Kadowaki1986}.
$T^2$ dependence of resistivity also appears 
in quasi-2-dimensional Fermi liquid of oxides such as Sr\sub{2}RuO\sub{4}\cite{Maeno1997},
which is strongly suggested as a spin-triplet superconductor with $T\subm{c}=1.5$~K.
Interestingly, a layered compound MgB\sub{2}, which shows superconductivity at 
$T\subm{c}=39$~K\cite{Nagamatsu2001},
also exhibits $T^2$ dependence in the resistivity up to 240~K\cite{Xu2001,Kim2002}.
The $\rho\propto T^2$ law in MgB\sub{2} is attributed not to the e-e correlation, 
but to the strong electron-phonon (e-ph) coupling 
between a quasi-2-dimensional conduction band 
and an optic phonon mode in the boron honeycomb layer\cite{Masui2002}.

\begin{figure}
\includegraphics[width=0.45\textwidth]{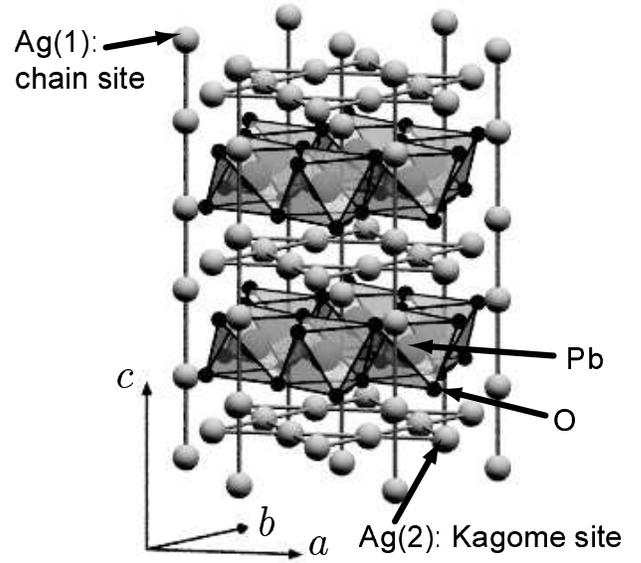}
\caption{Crystal structure of Ag$_5$Pb$_2$O$_6$.
\label{fig:crystal_structure}
}
\end{figure}

As another example of a layered system which exhibits $T^2$ resistivity,
but in a much wider temperature range \textit{up to room temperature},
we report here the physical properties of a layered oxide \apo.
This material was first reported by A.~Bystr\"{o}m and L.~Evers in 1950\cite{Bystrom1950}.
As shown in Fig.~\ref{fig:crystal_structure},
the crystal structure consists of an alternating stacking of Kagome layers of silver and 
honeycomb-like networks of PbO\sub{6} octahedra, 
which is connected by one-dimensional chains of silver penetrating the layers.
As a whole, \apo\ forms a hexagonal $P\bar{3}1m$ structure
with lattice parameters $a=b=5.9324(3)$\,\AA\
and $c=6.4105(4)$\,\AA\ \cite{Jansen1990}. 

Far before the report of the $\rho\propto T^2$ law,
it was suggested by B.~Dickens that \apo\ shows metallic conductivity\cite{Dickens1966},
and it was confirmed by M.~Jansen \etal\ in polycrystalline samples\cite{Jansen1990}.
Band calculation by T.~D.~Brennan \etal\ revealed 
that its Fermi level is located near the center of the band
and that its conductivity is attributed to one free electron from one formula unit, 
de-localized to all over the silver substructure; 
thus its valence formula is represented as 
(Ag$_5)^{4+}$(Pb$^{4+}$)\sub{2}(O$^{2-}$)\sub{6}\cite{Brennan1993}.
The $T^2$ dependence of the resistivity up to room temperature 
for the out-of-plane resistivity of single crystals 
was recently reported by Shibata \etal\ at meetings of the Physical Society of Japan\cite{ShibataJPS}.
This dependence is rather unusual, because resistivity of almost all metals show
$T$-linear dependence around room temperature due to the dominance of e-ph scattering.

It was Jansen \etal\ who first succeeded in controlled synthesis of polycrystalline \apo\ 
by high oxygen pressure method\cite{Jansen1990}.
Recent report by H.~Abe \etal\ on a new method to obtain single crystals\cite{Abe2002},
made it possible to perform more precise experiments.
Nevertheless, so far no article on experiments with single crystals has been published, 
although there are several articles on closely related materials,
such as Bi and In substitution to the Pb site\cite{Bortz1993}, 
Cu substitution to the Pb site\cite{Tejada2002} and 
\apo\ with defects\cite{Iwasaki2002}.
We note that there are also articles on powder samples by
D.~Djurek \etal\ reporting several strange properties\cite{Djurek2004,Djurek2004-02}, 
which have not been reproduced by other groups to the best of our knowledge.

We present here the first report on physical properties of single crystalline \apo .
We synthesized the single crystals and measured resistivity,
specific heat and magnetic susceptibility.
In resistivity, we succeeded not only in reproducing the $T^2$ dependence 
of the out-of-plane resistivity, but also in finding the same dependence of
the in-plane resistivity for the first time.
We will discuss the possible origin of the unusual $T^2$ resistivity.
What is more, we found a large signal of diamagnetic phase transition around 48~mK,
which implies that bulk superconducting phase exists in \apo.

\section{Experiments}

The single crystals of \apo\ used in this study were grown 
 by basically the same method as the one Shibata \etal\ reported\cite{ShibataJPS}.
As starting materials, we used AgNO$_3$ (99.9999\%)
 and Pb(NO$_3$)$_2$ (99.999\%).
We ground them in a mortar, 
 and put the powder into thin Al$_2$O$_3$ crusibles with a 10-mm inner diameter.
The ratio of Ag and Pb is basically stoichiometric.
The crusibles were heated in air to 80\degc\ and kept for 3 hours,
 then heated up to 300\degc\ in 8 hours 
 and up to the reaction temperature at 394\degc\ in another 5 hours.
We note here that one should cover the crusibles with larger crusibles or something
 in order to avoid pollution of the furnace, 
 since melting AgNO$_3$ sometimes splashes out of the crusibles.
The temperature was kept at this temperature for 48 hours, 
 and the crusibles were furnace-cooled to room temperature\cite{ShibataJPS}.

\begin{figure}
\includegraphics[width=0.45\textwidth]{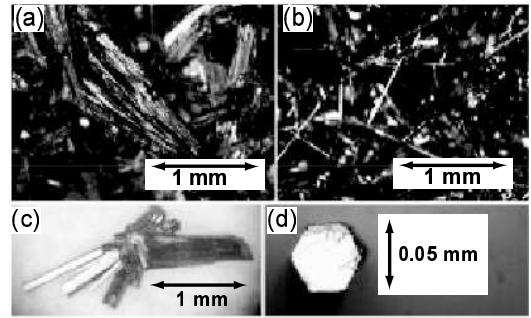}
\caption{Photos of single crystals of \apo\ taken with optical microscopes.
The long axis of these crystals are along the $c$ axis.
(a) Single crystals in a crusible which are obtained with 
 basic reaction parameters described in the text.
(b) Single crystals at the bottom of a crusible when we mixed AgNO$_3$
 and Pb(NO$_3$)$_2$ with a ratio of Ag:Pb=5:1 and also chose
 a lower reaction temperature (380\degc).
 The crystals grew dispersed from each other 
 and the shapes became better with the aid of self-flux AgNO$_3$.
 In this case AgNO$_3$ little remained since most of unreacted AgNO$_3$ had been evaporated
 or resolved into Ag metal.
(c) A group of single crystals removed from the crusible in photo (b).
 We obtained a lot of crystals with a perfectly hexagonal-stick shape 
 by adding self-flux AgNO$_3$.
(d) A cross section of a typical single crystal (not the largest). 
 This hexagonal plane is in the $ab$ plane of \apo.
\label{fig:photo_crystal}}
\end{figure}

However, the crystals obtained with the above parameters 
 do not have well-defined shapes as shown in Fig.~\ref{fig:photo_crystal}(a)
 thus they are not suitable for resistivity measurements.
We found that we can grow better-shaped crystals 
 by choosing lower reaction temperatures.
Furthermore, we also discovered that adding extra AgNO$_3$ promotes the growth,
 because AgNO\sub{3}, which melts at 212\degc, apparently plays a role of self-flux.
Figure~\ref{fig:photo_crystal}(b) shows the crystals grown with twice amount of AgNO$_3$
 and with a reaction at 380\degc.
Using this process,
 we obtained crystals in the shape of hexagonal 
 sticks, which are shown in Fig.~\ref{fig:photo_crystal}(c)(d).
The long axis of the sticks are along the $c$ axis.
The largest size of crystals was approximately 1~mm in length and 0.5~mm in thickness.

For characterization of the obtained crystals,
we used powder X-ray diffractmetry
with CuK$\alpha_1$ radiation, which confirmed that these crystals were single-phased \apo.
Resistivity was measured using a standard four-probe method with dc or ac current.
Specific heat was measured with a commercial calorimeter (Quantum Design, model PPMS),
and dc susceptibility by a SQUID magnetometer (Quantum Design, model MPMS).
AC susceptibility was measured by a mutual inductance method down to 38~mK 
using a {}$^4$He-{}$^3$He dilution refrigerator.

We note here about the lowest temperature of the resistivity measurement.
We used an ultrasonic soldering method to attach the leads to the crystals, 
because several conductive pastes (e.g. silver paste) we used all
damaged the sample seriously.
On the other hand, 
the solder we used (Cerasolzer \#123, Senju Metal Industry) becomes superconducting below 3~K, 
which makes it impossible for us to measure the sample resistivity accurately below 3~K\@.
We should also note that the accuracy of in-plane resistivity is
less reliable than that of the out-of-plane resistivity.
This is because we could not shape up the samples before attaching leads to the crystal.
Thus the electric current did not flow uniformly and we could not determine $\rho$
by the simple relation as $\rho=R(S/l)$, 
where $S$ is the cross section and $l$ is the lead spacing. 
The results of $\rho_{ab}$ shown in Fig.~\ref{fig:resistivity} is
an estimation assuming uniform current flow.
Nevertheless, the relative variation of $\rho$ in temperature should be reliable.

\section{Results}

\begin{figure}
\includegraphics[width=0.45\textwidth]{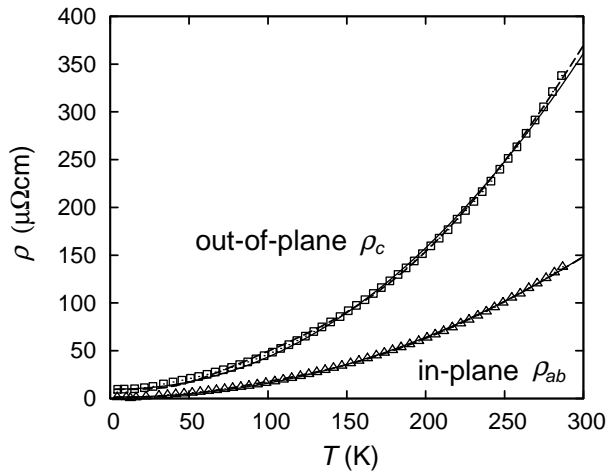}
\caption{Temperature dependence of the out-of-plane (open squares)
and the in-plane resistivity (open triangles) down to 3 K.
Solid lines are the fitting with $\rho(T)=\rho_0+A'T^p$ with $p=2.13$ for $\rho_c$
and $p=2.11$ for $\rho_{ab}$.
Dashed lines are the fitting assuming the existence of a Einstein-like optic phonon mode.
\label{fig:resistivity}}
\end{figure}

Figure~\ref{fig:resistivity} shows the resistivity 
along the $c$ axis ($\rho_c$) and in the $ab$ plane ($\rho_{ab}$).
Anisotropy of resistivity at 280~K is rather small and approximately 2,
which is consistent with the results of the band calculation by Brennan \etal\cite{Brennan1993}
The residual resistivity for $\rho_{c}$ and $\rho_{ab}$ are
9.7~$\mu\Omega$cm and 1.5~$\mu\Omega$cm, respectively;
this means residual resistivity ratio (RRR) are 35 for $\rho_c$
and 93 for $\rho_{ab}$, which show that the quality of the samples was quite good.
If we assume its temperature dependence as $\rho(T)=\rho_0+A'T^p$,
we obtained the exponent as $p=2.13$ for $\rho_c$ and $p=2.11$ for $\rho_{ab}$ 
from the fitting in all temperature range\cite{Remark_resistivity}.
For both direction, $p$ is close to 2;
thus if we approximate as $\rho(T)=\rho_0+AT^2$,
we obtained the coefficient of $T^2$ term 
$A_{c}=3.78\times 10^{-3}\:\mu\Omega$cm/K$^2$ for $\rho_c(T)$ and
$A_{ab}=1.6\times 10^{-3}\:\mu\Omega$cm/K$^2$ for $\rho_{ab}(T)$.

\begin{figure}
\includegraphics[width=0.45\textwidth]{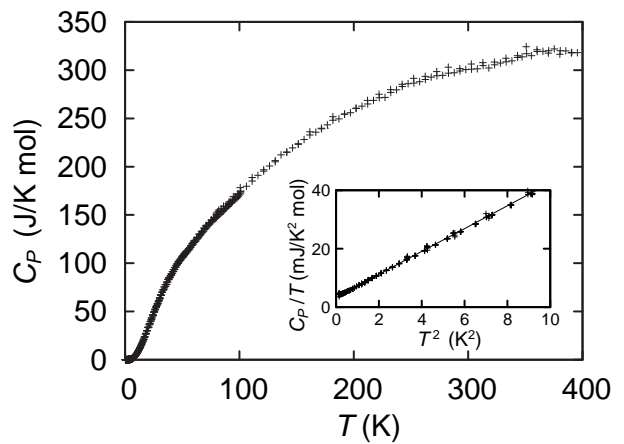}
\caption{Specific heat of Ag$_5$Pb$_2$O$_6$ down to 0.35~K\@.
These figures contain the results of several different samples.
Inset: $C_P/T$ plotted against $T^2$ below 3.3~K\@. 
Solid line is the fitting $C_P(T)/T=\gamma\subm{e}+\beta\subm{p}T^2$ with
$\gamma\subm{e}=3.42$~mJ/K$^2\:$mol and 
$\beta\subm{p}=3.91$~mJ/K$^4\:$mol. 
\label{fig:heatcapacity}
}
\end{figure}

The measured specific heat is shown in Fig.~\ref{fig:heatcapacity}.
Since \apo\ does not contain any magnetic ions, we express $C_P$ in 1-mol formula unit,
 in other words, of 1-mol conduction electrons.
In high temperature region, $C_P$ converges to a constant value,
 which agrees with Dulong-Petit's classical phonon specific heat,
 $C\subm{cl}=3k\subm{B}rN\subm{A}=324$~J/K\,mol\cite{Remark_heatcapacity}, 
 where $r=13$ is the number of atoms in one formula unit.
We also measured the specific heat in magnetic fields, 
 but there was no difference between the data in 0 T and in 7 T\@.
The inset is the $C_P/T$-vs.-$T^2$ plot below 3.3~K\@.
We obtained the electronic specific heat coefficient $\gamma\subm{e}=3.42$~mJ/K$^2\:$mol
 and the Debye temperature $\mathit{\Theta}\subm{D}=186$~K from this plot.

\apo\ displays weak constant diamagnetism in $10\text{~K}<T<350\text{~K}$, 
$(-2.07\pm 0.02)\times 10^{-4}$~emu/mol, from dc susceptibility measurement.
Below 10~K, $\chi\subm{dc}$ increases probably owing to paramagnetic impurities.
To subtract the diamagnetic contribution from the ion cores, 
we used the values of the susceptibility of cores in the literature\cite{MagnetoChem};
we estimated the total contribution from the cores as 
$-2.44\times 10^{-4}$~emu/mol.	
When we subtracted this value from the experimental value, 
 we obtained $(+3.7\pm0.2) \times 10^{-5}$~emu/mol.
This is the Pauli susceptibility of conduction electrons
 because there are no magnetic ions in \apo.

\begin{figure}
\includegraphics[width=0.45\textwidth]{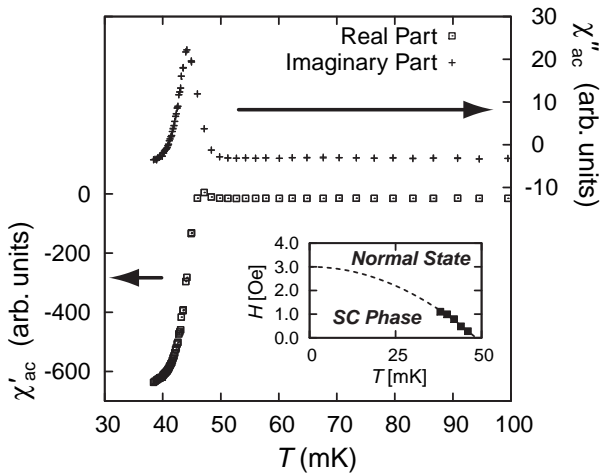}
\caption{AC magnetic susceptibility below 0.1~K,
measured with an ac magnetic field of amplitude $2.3\times10^{-2}$~Oe and frequency 887~Hz.
A very large diamagnetic phase transition signal was found below 45~mK,
which suggests the existence of a superconducting (SC) phase.
Inset: 
Phase diagram of \apo\ below 50~mK.
Closed squares are ``$H\subm{c}$'' determined from the peaks of 
imaginary part $\chi ''\subm{ac}$.
Dashed line shows the classical parabolic law of superconductivity:
$H\subm{c}(T)=H\subm{c0}[1-(T/T\subm{c0})^2]$.
\label{fig:ac_susceptibility}
}
\end{figure}

In the ac susceptibility measurement, we found a very large signal around 45~mK
 as shown in Fig.~\ref{fig:ac_susceptibility}.
The signal was reproducible in different samples, in different susceptibility cells
 and in different ac magnetic field amplitudes and frequencies.
We also measured ac susceptibility of a bulk pure Al with approximately the same volume
 as the measured \apo\ using the same cell, and 
 found that \apo\ yields as large a diamagnetic signal 
 as that of the superconducting transition of Al.
Moreover, this transition disappears under a few~Oe magnetic field, and
 \apo\ does not contain any magnetic ions which may cause a diamagnetic transition.
Thus this signal strongly implies the existence of a bulk superconducting phase of \apo.
The inset of Fig.~\ref{fig:ac_susceptibility} is the phase diagram of this new phase.
 ``$H\subm{c}(T)$'' was determined from the peaks of 
 imaginary part $\chi ''\subm{ac}$ by sweeping magnetic field at each fixed temperature.
We note that values of $H$ in the inset of Fig.~\ref{fig:ac_susceptibility} is shifted 
because exact zero point of our magnet changed by a few Oe
due to the residual magnetic field.
Thus we swept magnetic field from $+10$~Oe to $-10$~Oe
and determined exact zero point from the symmetry of the $\chi\subm{ac}$ data.
``$H\subm{c}$'' is well fitted by a parabolic law for superconductor: 
$H\subm{c}(T)=H\subm{c0}[1-(T/T\subm{c0})^2]$ 
where $H\subm{c0}=3.0$~Oe and $T\subm{c0}=48$~mK obtained by fitting. 
This variation also supports the evidence of superconductivity.

\section{Discussion}

Now we discuss the origin of the $T^2$ dependence of resistivity.
One possibility is e-e interaction.
Resistivity of materials in which strong e-e correlation exists 
 shows $T^2$ dependence only at temperature much lower than room temperature,
 because the characteristic temperature of many-body Fermi liquid formation is low
 and also the $T^2$-behavior requires the dominance 
 of the e-e scattering over the e-ph scattering.
If the $\rho\propto T^2$ law up to \textit{room temperature} of \apo\ 
comes from the e-e correlation,
the e-e correlation must be so large as to dominate over the e-ph scattering
even at room temperature.
But it seems impossible to expect strong e-e interaction from the discussions below.
First, the electronic specific heat coefficient of \apo\ 
is comparable to those of normal metals and 1 $\sim$ 3 orders of magnitude smaller than
those of strongly-correlated electron systems.
Moreover, neither the Wilson ratio nor the Kadowaki-Woods (K-W) ratio
shows evidence of strong e-e correlations.
The Wilson ratio, which is given by 
$(\pi^2k\subm{B}^2/3\mu\subm{B}^2)(\chi\subm{Pauli}/\gamma\subm{e})$,
is expected to increase from unity to 2 
as e-e interaction increases from 0 to $\infty$\cite{Yamada1975}.
In \apo, the Wilson ratio is 0.79, close to unity, 
implying that the e-e interaction is weak.
Of course we should note that the value of $\chi\subm{Pauli}$ may not be very accurate
because the subtracted core contribution is much larger than $\chi\subm{Pauli}$.
The K-W ratio is defined as $A/\gamma\subm{e}^2$ and becomes a universal value
$a_0=1.0\times 10^{-5}\ \mu\Omega$cm(K\,mol/mJ)$^2$ 
in most strongly-correlated systems\cite{Kadowaki1986,Miyake1989}.
The K-W ratio of \apo\ is 
$32.3a_0$ for $\rho_c$
and approximately $14a_0$ for $\rho_{ab}$; both are much larger than $a_0$.
These deviation imply that \apo\ cannot be classified into strongly-correlated electron systems
and that the $T^2$ dependence of $\rho$ is not due to e-e interaction.

Another possible origin is the e-ph scattering.
Basically, resistivity governed by the e-ph interaction is $T$-linear above 
 $\mathit{\Theta\subm{D}}/5$ ($\approx 20\sim 100$~K, in most metals).
But there are some materials
 of which resistivity shows non-linear $T$ dependence  
 at relatively high temperature because of the  e-ph scattering by high energy optic phonons.
For example, MgB\sub{2} shows $\rho\propto T^2$ behavior 
 in $T\subm{c}<T\lesssim 240\text{~K}$\cite{Xu2001,Kim2002},
 and ReO\sub{3} exhibits an unusual curvature in the resistivity 
 in $120\text{~K}\lesssim T\lesssim 220\text{~K}$\cite{King1971}. 
When analyzing non-linear $T$ dependence of resistivity in these materials,
 it is a common approach to assume the optic phonons as 
 Einstein-like\cite{King1971,Allen1993,Masui2002}.
In this method, the Boltzmann equation is a starting point:
%%%%%%
\begin{align}
\rho(T)= 
\rho_0 + \frac{(4\pi)^2k\subm{B}T}{\hbar\omega\subm{p}^2}%
\!\!\int^{\omega\subm{max}}_0\!\frac{d{\omega}}{\omega}\alpha^2F({\omega})
\left(\frac{x}{\sinh x}\right)^{\!\!2}, \label{eq:Boltzmann}
\end{align}
%%%%%%
 where $x=\hbar\omega/2k\subm{B}T$, $\alpha^2F(\omega)$ is the Eliashberg function, 
 $\omega\subm{p}$ is the plasma frequency of conduction electrons.
We assume that we can separate the Eliashberg function 
 into the Debye term and the Einstein term:
%%%%%%
\begin{align}
 \alpha^2F(\omega)= 
 2\lambda\subm{D}\!\left(\!\frac{\omega}{{\Omega\subm{D}}}\!\right)^{\!\!4}
 \!\theta(\Omega\subm{D}\!-\!\omega)+
 \frac{\lambda\subm{E}\Omega\subm{E}}{2}\delta(\omega\!-\!\Omega\subm{E}),
 \label{eq:Eliashberg}
\end{align}
%%%%%%
 where $\Omega\subm{D}$ and $\Omega\subm{E}$ are 
 the Debye frequency and the frequency of Einstein phonons, 
 respectively, and $\lambda\subm{D}$ and $\lambda\subm{E}$ are 
 the e-ph coupling constant of Debye phonons and of Einstein phonons, respectively.
Now we can substitute this Eliashberg function into Eq.~\ref{eq:Boltzmann}, 
 and fit it with the experimental data.
When we used $k\subm{B}\Omega\subm{D}/\hbar$, $k\subm{B}\Omega\subm{E}/\hbar$, 
 $\lambda\subm{D}/\omega\subm{p}^2$
 and $\lambda\subm{E}/\omega\subm{p}^2$ as
 fitting parameters, 
we obtained 
$k\subm{B}\Omega\subm{D}/\hbar=290$~K and $k\subm{B}\Omega\subm{E}/\hbar=1200$~K 
 from the fitting with $\rho_c$ data,
 and $k\subm{B}\Omega\subm{D}/\hbar=280$~K and $k\subm{B}\Omega\subm{E}/\hbar=1100$~K 
 from $\rho_{ab}$.
 The results of the fittings are shown in Fig.~\ref{fig:resistivity} with dashed lines.
But the existence of such a high frequency Einstein-like phonon is doubtful.
This is because specific heat shown in Fig.~\ref{fig:heatcapacity} converges to
 Dulong-Petit's value around 350~K,
 implying that all phonon modes are ``freely'' excited at this temperature
 and thus there seems to be no more phonon mode 
 with energy much larger than 350~K\cite{Remark_Raman}.
For comparison, the specific heat of MgB\sub{2}, 
 in which the optic phonon mode with a wavenumber of 620~cm$^{-1}$ (890~K in energy)
 plays important roles,
 is indeed not at all saturated even at 300~K\cite{Wang2001}.

If we assume that the novel phase below 48~mK is a phonon-mediated superconducting phase, 
 we can estimate the e-ph coupling constant $\lambda$ 
 from McMillan's relation\cite{McMillan1968}:
%%%%%%%%%%%%%%%%%%%%%%
\begin{align}
 T\subm{c}=\frac{\mathit{\Theta}\subm{D}}{1.45}%
 \exp\left[-\frac{1.04(1+\lambda)}{\lambda-\mu^\ast(1+0.62\lambda)} \right].
\end{align}
%%%%%%%%%%%%%%%%%%%%%%
When we assumed the psudopotential as $\mu^\ast=0.1$ 
 and used $\mathit{\Theta}\subm{D}=186$~K, 
 we obtained $\lambda=0.288$, which is comparable to those of 
 W ($\lambda=0.28,\:T\subm{c}=12$~mK) or
 Be ($\lambda=0.23,\:T\subm{c}=26$~mK)\cite{McMillan1968}.
Such a small $\lambda$ may be another evidence 
 that strong e-ph interaction does not exist in \apo.

The unusual $\rho\propto T^2$ behavior up to room temperature thus can be attributed to
 neither e-e interaction nor e-ph interaction including the optic phonons.
We note here that a one-dimensional chain organic compound TTF-TCNQ
 exhibits $T^{2.33}$ dependence of resistivity in $60\text{~K}<T<300\text{~K}$\cite{Groff1974},
 attributable to the librational phonon mode in the chain\cite{Gutfreund1977}.
This scenario is not applicable to \apo\ because of 
 the rigidness of the lattice against twisting motion of the silver chain.
Nevertheless, a unique phonon mode may exist also in \apo\ 
 which gives rise to the $T^2$ dependence.
Thus a detailed investigation into the phonon modes in \apo\
 and into their strength of coupling with conduction electrons 
 may elucidate the $T^2$ dependence.

\section{Conclusion}

In conclusion, we synthesized single crystals of \apo\ and 
 measured their resistivity, specific heat and magnetic susceptibility.
The resistivity exhibits unusual $T^2$ dependence up to room temperature
 both along the $c$ axis and in the $ab$ plane.
This behavior cannot be explained by e-e interaction.
It seems also impossible to explain the $T^2$ dependence from 
 the e-ph scattering by optic phonons.
Thus there should be other scattering mechanisms in \apo.
Moreover, a phase transition with a large diamagnetic signal was found 
 below 48~mK and below 3~Oe.
This new phase seems to be a bulk superconducting phase in \apo.

\acknowledgements

We would like to thank Prof.~Y.~Yamada at Niigata University,
K.~Deguchi and K.~Kitagawa at Kyoto University for their great help for measurements and 
their useful discussions.
We also acknowledge Prof.~K.~Tanaka and Y.~Okuda 
at Kyoto University for the Raman spectroscopy
and S.~Fujimoto at Kyoto University for his helpful suggestions.
This work is supported by a Grant-in-Aid for the 21st Century COE 
``Center for Diversity and Universality in Physics'' from MEXT of Japan.
It is also in part supported by additional Grants-in-Aid from JSPS and MEXT.

\bibliography{Ag5Pb2O6}

\begin{thebibliography}{29}
\bibitem[{Rem({\natexlab{b}})}]{Remark_heatcapacity}
\bibinfo{note}{Strictly speaking, {$C_{\text{cl}}$} is the specific heat at
  constant volume while {$C_P$} is at constant pressure; however the difference
  of them should be small.}
\end{thebibliography}

\end{document}